\documentclass[preprint,3p,authoryear,a4paper]{elsarticle}
\usepackage{graphicx}
\usepackage{slashbox}
\usepackage{color}
\usepackage{dsfont}
\usepackage{amsmath}
\usepackage{amssymb}
\usepackage{amsfonts}
\usepackage{amsbsy}
\usepackage{caption}
\usepackage{subcaption}
\usepackage{amsthm}
\usepackage{array}
\usepackage{booktabs} 
\usepackage{verbatim}
\usepackage[english]{babel}
\usepackage{natbib}
\usepackage{ulem}
\usepackage{url}
\usepackage{multirow}
\usepackage{subcaption}
\usepackage{float}
\usepackage{numprint}
\usepackage{enumitem}
\usepackage[title]{appendix}
\usepackage[font=normalsize, labelfont=bf]{caption}
\usepackage{tabularx}
\usepackage{ltablex}
\usepackage{makecell}
\usepackage[table,xcdraw]{xcolor}

\numberwithin{table}{section}

\newcolumntype{R}{>{\raggedleft\arraybackslash}X} % right align fixed width columns
\newcolumntype{P}[1]{>{\centering\arraybackslash}p{#1}} % centered fixed width column

\renewcommand\appendix{\par
\setcounter{section}{0}
\setcounter{subsection}{0}
\setcounter{table}{0}
\setcounter{figure}{0}
\gdef\thetable{\Alph{table}}
\gdef\thefigure{\Alph{figure}}
\gdef\thesection{\Alph{section}}
\setcounter{section}{0}}

\numberwithin{equation}{section}

\newcounter{arclist}

\newcounter{arcenum}

\newcommand{\rev}[1]{\textcolor{black}{#1}}
\newcommand{\reva}[1]{\textcolor{black}{#1}}

\begin{document}

\normalem

\begin{frontmatter}

\title{\texttt{SynthETIC}: an individual insurance claim simulator with feature control}

\author[UMelb]{Benjamin Avanzi}
\ead{b.avanzi@unimelb.edu.au}

\author[UNSW]{Greg Taylor\corref{cor}}
\ead{gregory.taylor@unsw.edu.au}

\author[UNSW]{Melantha Wang}
\ead{chenyi.wang@student.unsw.edu.au}

\author[UNSW]{Bernard Wong}
\ead{bernard.wong@unsw.edu.au}

\cortext[cor]{Corresponding author. }

\address[UMelb]{Centre for Actuarial Studies, Department of Economics, University of Melbourne VIC 3010, Australia}
\address[UNSW]{School of Risk and Actuarial Studies, UNSW Australia Business School, UNSW Sydney NSW 2052, Australia}

\begin{abstract}
Recent years have seen rapid increase in the application of machine learning to insurance loss reserving. They yield most value when applied to large data sets, such as individual claims, or large claim triangles. In short, they are likely to be useful in the analysis of any data set whose volume is sufficient to obscure a naked-eye view of its features. Unfortunately, such large data sets are in short supply in the actuarial literature. Accordingly, one needs to turn to synthetic data. Although the ultimate objective of these methods is application to real data, the use of synthetic data containing features commonly observed in real data is also to be encouraged. 

While there are a number of claims simulators in existence, each valuable within its own context, the inclusion of a number of desirable (but complicated) data features requires further development. Accordingly, in this paper we review those desirable features, and propose a new simulator of individual claim experience called \texttt{SynthETIC}. 

Our simulator is publicly available, open source, and fills a gap in the non-life actuarial toolkit. The simulator specifically allows for desirable (but optionally complicated) data features typically occurring in practice, such as variations in rates of settlements and development patterns; as with superimposed inflation, and various discontinuities, and also enables various dependencies between variables. The user has full control of the mechanics of the evolution of an individual claim. As a result, the complexity of the data set generated (meaning the level of difficulty of analysis) may be dialled anywhere from extremely simple to extremely complex. The default version is parameterized so as to include a broad (though not numerically precise) resemblance to the major features of experience of a specific (but anonymous) Auto Bodily Injury portfolio, but the general structure is suitable for most lines of business, with some amendment of modules.

\end{abstract}

\end{frontmatter}
{\centering \large}

\section{Introduction}\label{sec:intro}
Recent years have seen rapid increase in the application of \textbf{machine learning (ML)} to insurance loss reserving. \rev{Examples of various ML techniques applied in aggregated loss reserving include, for example, \citet{EnVe01} who applied GAM, \citet{ClSpSc14} who considered GAMLSS, \citet{McTaMi18} who applied LASSO techniques. Neural-network based approaches were also considered by  \citet{Wu18_nn}, \citet{Ku19}, and \citet{Ga20}. Similarly, ML techniques have also been used in models of individual claims reserving, including for example \citet{Wu18} who applied tree based methods, and \citet{Mu06}, \citet{DeLiWu20} who applied neural-network based models. See also  \citet{Ri20a, Ri20b}, and \citet{BlCoLaMa20} for additional commentary and extensive references.} 

These ML methods are hungry for data. They yield most value when applied to large data sets, such as individual claim, even transactional, data sets. They may also be useful in the analysis of large claim triangles, e.g. $40 \times 40$. In short, they are likely to be useful in the analysis of any data set whose volume is sufficient to obscure a naked-eye view of its features.

Unfortunately, such large data sets are in short supply in the actuarial literature, and so there is a shortfall in the material required to test ML methods. Accordingly, one needs to turn to synthetic data. Although the ultimate objective of these methods is application to real data, the use of synthetic data containing features commonly observed in real data is also to be encouraged.

Indeed, there is a school of thought that proposed claim models should \textbf{always} be tested against synthetic data, as well as real data if available. The reason is that the features contained in the synthetic data set will be known, by construction. One is then able to test the extent to which the proposed model is successful in identifying them.

There are several individual claim simulators in existence, and they are briefly reviewed in Section~\ref{sec:claimsim_lit}. Although each of these is valuable within its own context, we believe that there is scope for a further simulator due to the need for inclusion of a number of desirable (but complicated) data features, see Section~\ref{sec:prior_lit}. It is against this background that the individual claim simulator \textbf{\texttt{SynthETIC}} (\textbf{S}ynthetic \textbf{E}xperience \textbf{T}racking \textbf{I}nsurance \textbf{C}laims) is introduced. It is publicly available, open source (see Section \ref{sec:sim_repository}), \rev{and is the only claims simulator available as a full \texttt{R} package on CRAN, with all the associated documentation.}. \rev{Furthermore, it is fully compatible with existing actuarial CRAN packages such as \texttt{actuar} \citep{R-actuar}}, and fills a gap in the non-life actuarial toolkit. 

\texttt{SynthETIC} simulates, for each claim, occurrence, notification, the timing and magnitude of individual partial payments, and closure. \rev{While \texttt{SynthETIC} comes with a default set of distributional assumptions, the flexibility of \texttt{SynthETIC} implies that users of the program are free to choose and implement any form of distribution models with ease.} Inflation, including (optionally) superimposed inflation, is incorporated in payments. Superimposed inflation may occur over calendar or accident periods. The \rev{micro-level} claim data \rev{can be} summarized by accident and payment ``periods'' whose duration is an arbitrary choice (e.g. month, quarter, etc.) available to the user, \rev{which is then ready for chain ladder reserving analysis by \texttt{ChainLadder} \citep{R-ChainLadder}.}

The code is structured as eight modules (occurrence, notification, \ldots), any one or more of which may be re-designed according to the user’s requirements. The default version is parameterized so as to include a broad (though not numerically precise) resemblance to the major features of experience of a specific (but anonymous) Auto Bodily Injury portfolio.  It thus reflects a number of desirable (but complicated) data features, but the general structure is suitable for most lines of business, with some amendment of modules. The structure of the simulator enables the inclusion of a number of important dependencies between the variables related to an individual claim, e.g. dependence of notification delay on claim size, of the size of a partial payment on the sizes of those preceding, etc. \rev{Many users should be able to use \texttt{SynthETIC} without modifying such dependence structure between individual claim features. However, \texttt{SynthETIC} is flexible enough to accommodate alternative dependencies. Ultimately,} the user has full control of the mechanics of the evolution of an individual claim. 

Thanks to the high flexibility of \texttt{SynthETIC}, the complexity of the data set generated (meaning the level of difficulty of analysis) \emph{may be dialled anywhere from extremely simple to extremely complex}. At the extremely simple end would be chain-ladder-compatible data, and so the alternative data structures available enable proposed loss reserving models to be tested against more challenging data sets. Indeed, the user may generate a collection of data sets that provide a spectrum of complexity. This is particularly useful when trying to `break' a model and ascertaining its qualities and limitations.

This paper is structured as follows. After a discussion of desirable features in Section~\ref{sec:prior_lit}, we present the notation and details of \texttt{SynthETIC} in the following sections. The claim process is defined by 8 modules (see Section~\ref{sec:claim_process}), any of which may be unplugged and replaced by an alternative module of the user’s choosing. The current default parameterization broadly resembles a specific real Auto Liability data set that contains a number of features that render its modelling demanding.

After some notation in Section~\ref{sec:notation}, the architecture of \texttt{SynthETIC} is described in Section~\ref{sec:architecture}. Section~\ref{sec:prior_lit} discusses its relation to prior literature, while Section~\ref{sec:ex_implementation} gives detail of the default parameterization just mentioned. Section~\ref{sec:application} discusses the circumstances and manner in which \texttt{SynthETIC} might be applied, Section~\ref{sec:sim_repository} gives the web address of the code and the example data set used in Section~\ref{sec:ex_implementation}, and Section~\ref{sec:conclusion} contains some closing comments.

\section{Desirable data features}
\label{sec:desirable features}
Much of the actuarial literature on loss reserving is focused on the chain ladder. This model requires a very specific parametric structure in which the expectation of the target variable is equal to the product of a row parameter and a column parameter \citep[see, for example,][]{Tay00,WuMe08}.

However, the same literature is dotted with examples of data sets that do not fit this prescription.  Indeed, various models within the literature have been devised with the express purpose of addressing such non-conforming data sets.

This failure to conform with the chain ladder structure can arise in a number of ways. First, the \textbf{rate of settlement of claims} can change from one accident period to another. Models that take account of this feature have a long pedigree, e.g. \citet{FiLa73,Rei78} through to the present day, e.g. \citet{McTaMi18}.

More generally, \textbf{development patterns} may change from one accident period to another. This may be due to changing rates of settlement, changing mixture of risks, or other causes. Again, models that seek to address this feature have a long pedigree, e.g. \citet{BeSh77} up to recent times, e.g. \citet{Mey15}.

\textbf{Superimposed inflation} is a further potential disruptor of claim experience. The chain ladder model can accommodate inflation that occurs at a constant rate over time, but not variable-rate inflation. Examples of real data sets that appear to exhibit variable rates of superimposed inflation appear in e.g. \citet{Tay00} and \citet{McTaMi18}.

A claim triangle may experience a \textbf{discontinuity} as one passes from one accident period to another. This might result from a change in legislation governing claim conditions, or a sudden change in the type of business underwritten. An example is discussed by \citet{McG07b}.

Models that have been created to address data features such as the above require data sets containing those features if they are to be tested effectively. A major objective of \texttt{SynthETIC} is to provide this facility, and to our knowledge, it is the only simulator (at the time of writing this paper) that includes all those features natively. 

\section{Notation} \label{sec:notation}
\texttt{SynthETIC} works with \textbf{exact transaction times}, so time will be measured continuously. Calendar time $\tau = 0$ denotes the first date on which there is exposure to occurrence of a claim. The time scale is arbitrary; a unit of time might be a quarter, a year, or any other selected period. The length of a period in years is specified by the user as a global parameter.  The user needs to ensure that all input parameters are compatible with the chosen time unit.

For certain purposes (see Section~\ref{sec:claim_process}), it will be useful to partition time into discrete periods. These are unit periods according to the chosen time scale. These periods will be of two types:
\begin{itemize}
    \item \textbf{occurrence periods} (or accident periods), numbered $1, 2, \dots$, where occurrence period $1$ corresponds to the calendar time interval $(0, 1]$;
    \item \textbf{payment periods}, numbered $1,2,\ldots,2I-1$ representing the calendar periods in which individual payments are made, and including $I$ past periods and a further $I-1$ future ones (i.e.  the final payment period corresponds to the final development period of the final occurrence period).
\end{itemize}
An individual claim is settled by means of one or more separate payments, referred to here as \textbf{partial payments}. The claim will be regarded as settled immediately after the final partial payment. The delays between successive partial payments are referred to as \textbf{inter-partial delays}.

All payments are subject to inflation. They are initially simulated without allowance for inflation, and an inflation adjustment added subsequently. Any quantity described as ``without allowance for inflation'' is expressed in constant dollar values, specifically those of calendar time $\tau = 0$.

Inflation occurs in two types:
\begin{enumerate}[label=(\alph*)]
    \item \textbf{Base inflation}: which represents, in some sense, ``normal'' community inflation (e.g. price inflation, wage inflation) that would apply to claim sizes in the absence of extraordinary considerations; and
    \item \textbf{Superimposed inflation}: which represents the differential (positive or negative) between claim inflation and base inflation.
\end{enumerate}

It is assumed that base inflation may be represented by a vector of quarterly inflation rates for both past and future calendar periods.  The input inflation rates need to be expressed as quarterly effective rates irrespective of the length of calendar periods adopted.  The inflation rates are used to construct an inflation index whose values are obtained:
\begin{itemize}
    \item at quarterly points from calendar time 0, by compounding the quarterly rates; and
    \item at intra-quarterly points, by exponential interpolation between the quarter ends immediately prior and subsequent.
\end{itemize}
It is also assumed that superimposed inflation occurs in two sub-types:
\begin{enumerate}[label=(\roman*)]
    \item \textbf{Payment period superimposed inflation}: which operates over payment periods; and
    \item \textbf{Occurrence period superimposed inflation}: which operates over occurrence periods.
\end{enumerate}
By default, both will be applied jointly, but this can be controlled by the user.

The following notation is used throughout:
\begin{itemize}[label=] \setlength\itemsep{0em}
    \item $\mathrm{Int}(x)$ denotes the integral part of $x$
    \item $\lceil x\rceil =$ the ceiling function, that is, the smallest integer $n$ such that $n-1\le x\le n$
    \item $i =$ occurrence period
    \item $\overline{t} =$ continuous calendar time with origin at the beginning of occurrence period 1
    \item $t =\lceil t\rceil$ payment period
    \item $E_i =$ (annual effective) exposure in occurrence period $i$
    \item $\lambda_i =$ expected claim frequency (per unit exposure) in occurrence period $i$
    \item $f(\overline{t}) =$ base inflation index, representing the ratio of dollar values at calendar time $\overline{t}$ to those at calendar time 0, constructed from the input base inflation rates
    \item $g_P(\overline{t} \mid s) =$ payment period superimposed inflation index, representing the ratio of dollar values at calendar time $\overline{t}$ to those at calendar time 0
    \item $g_O(i \mid s) =$ occurrence period superimposed inflation index, representing the ratio of dollar values at occurrence period $u$ to those at occurrence time 0
    \item $n_i =$ number of claims occurring in occurrence period $i$
    \item $r =$ identification number of claims occurring in occurrence period $i$ ($r = 1, 2, \dots, N_i)$
    \item $u_{ir} =$ occurrence time of claim $r$ of occurrence period $i$
    \item $s_{ir} =$ size of claim $r$ of occurrence period $i$ without allowance for inflation
    \item $v_{ir} =$ delay from occurrence to notification of claim $r$ of occurrence period $i$ (N.B. the notification time is $u_{ir} + v_{ir}$)
    \item $w_{ir} =$ delay from notification to settlement of claim $r$ of occurrence period $i$ (N.B. the settlement time is $u_{ir} + v_{ir} + w_{ir}$)
    \item $m_{ir} =$ number of partial payments in respect of claim $r$ of occurrence period $i$
    \item $s_{ir}^{(m)} =$ size of the $m$-th partial payment in respect of claim $r$ of occurrence period $i$, $m = 1, 2, \dots, m_{ir}$
    \item $p_{ir}^{(m)} = s_{ir}^{(m)} / s_{ir} =$ proportion of claim amount $s_{ir}$ paid in the $m$-th partial payment
    \item $d_{ir}^{(m)} =$ the inter-partial delay from the epoch of the $(m-1)$-th to the $m$-th partial payment of claim $r$ of occurrence period $i$, with the convention that $d_{ir}^{(0)} = 0$, corresponding to notification date (the epoch of the $m$-th partial payment is $u_{ir} + v_{ir} + d_{ir}^{(1)} + \dots + d_{ir}^{(m)}$)
\end{itemize}
All of these quantities from $n_i$ onward, but except $r$, are realizations of random variables. The random variables themselves are denoted in the same way but with the primary symbol in upper case. For example, $S_{ir}$ denotes the random variable whose realization is $s_{ir}$.

\section{Architecture} \label{sec:architecture}

\subsection{Claim process} \label{sec:claim_process}
The claim process for claim $r$ of occurrence period $i$ is envisaged as consisting of the following modules:
\begin{itemize}
    \item \ref{sec:module1} -- Claim occurrence date; 
    \item \ref{sec:module2} -- Claim size without allowance for inflation; 
    \item \ref{sec:module3} -- Claim notification date; 
    \item \ref{sec:module4} -- Claim settlement date; 
    \item \ref{sec:module5} -- Number of partial payments; 
    \item \ref{sec:module6} -- Sizes of partial payments without allowance for inflation; 
    \item \ref{sec:module7} -- Distribution of payments over time; 
    \item \ref{sec:module8} -- Claim inflation. 
\end{itemize}
The modules have been listed here in the logical order of \texttt{SynthETIC} rather than temporal order. This architecture more or less follows that set out in the early papers on claim micro-models \citep{Arj89, Jew89, Nor93, Nor99, Hes94}. The following paragraphs provide full detail of the architecture of each module, though each may be re-coded by the user if desired (see Section~\ref{sec:modularity}).

\rev{The \textbf{default version} of the package encodes the modules described in the remainder of the present sub-section, in conjunction with the parameterization set out in Appendix~\ref{app:parameterizations}.}

\subsubsection{Claim occurrence date} \label{sec:module1}
Occurrence period $i$ covers calendar period $(i-1,i]$. It is assumed \rev{by default} that $N_i \sim \mathrm{Poisson}(E_i \lambda_i)$, where $E_i, \lambda_i$ are input parameters. The number of claims $n_i$ is a random drawing of $N_i$. As indicated in Section \ref{sec:notation}, $E_i$ is an \textbf{annual effective} parameter. \rev{In addition to the default Poisson frequency, \texttt{SynthETIC} supports a full range of user-specified alternative sampling distributions, from standard parametric ones (e.g. negative binomial) to more customized versions (e.g. non-homogeneous Poisson process); as is the case for all other modules that follow.}

\rev{\texttt{SynthETIC} is also fully compatible with existing actuarial \texttt{R} packages such as \texttt{actuar} or \texttt{fitdistrplus}, which means that users can always pick a more advanced (and also more realistic) claim frequency or severity distribution introduced by \texttt{actuar}, or simulate from a parametric model fitted on a real data set by \texttt{fitdistrplus} \citep{R-actuar,R-fitdistrplus}. Examples are documented in the package vignette (see Section~\ref{sec:sim_repository}).} The example implementation in Section \ref{sec:ex_implementation} is specified in Appendix \ref{app:parameterizations} to work with quarterly periods, and with $E_i=12000$, $\lambda_i=0.03$.  The exposure for a single occurrence quarter will therefore be 3000 (exposure-years), and the expected claim frequency $3000\times 0.03=90$. 

No trend in claim frequency over the occurrence period has been assumed, and so the occurrence time of claim $r$ is a realization $u_{ir}$ of a random variable $U_{ir}$ simulated as $U_{ir} \sim \mathrm{Uniform}(i-1,i]$, $r= 1,2, \dots, N_i$.

In principle, the assumption of uniformity would conflict with fractional $E_i$ if the position of these exposures within the occurrence period were known. This conflict has been disregarded for the purpose of the \rev{default} version of \texttt{SynthETIC}, but is capable of modification if the conflict is material.

The \rev{default} selections of $E_i, \lambda_i$ are set out in Appendix~\ref{app:parameterizations}.

\subsubsection{Claim size without allowance for inflation} \label{sec:module2}
Claim size is represented as a multiple of a reference claim size, which is defined as a global parameter.  This enables the simulator to switch conveniently between currencies, e.g. the reference claim size might be 1,000 USD or 1,000,000 KRW.  Alternatively, some schemes of insurance define entitlements as multiples of a regulated reference claim size, and this latter may be used as the \texttt{SynthETIC} global parameter.

Claim size $s_{ir}$ is the realization of a random variable $S_{ir}$ with df $F_S(s)$, independent of $i$ in this version of \texttt{SynthETIC}, specified on input as a function of $s$. This is sampled according to $S_{ir} = F_S^{-1} \left( Z_{ir}^{(S)} \right)$, where $Z_{ir}^{(S)} \sim \mathrm{Uniform}(0,1]$. \rev{In practice, the user can choose to specify to \texttt{SynthETIC} either a distribution function (based on which \texttt{SynthETIC} applies inverse transform sampling), or directly a random generating function. The flexibility with the latter implies that the user is able to simulate from almost any model. The \texttt{SynthETIC} vignette includes an example on simulating claim sizes from a gamma GLM, and many others on alternative parametric models.}

The \rev{default} version of $F_S(s)$ is set out in Appendix~\ref{app:parameterizations}.

\subsubsection{Claim notification date} \label{sec:module3}
Notification delay $v_{ir}$ is the realization of a random variable $V_{ir}$ with df $F_{V|i,s}(v;i,s)$, specified on input as a function of $v$, and possibly dependent on occurrence period $i$ and claim size $s$. This is sampled according to $V_{ir} = F_{V|i,s}^{-1} \left(Z_{ir}^{(V)} \right)$, where $Z_{ir}^{(V)} \sim \mathrm{Uniform}(0,1]$. \rev{Here, $F_{V|i,s}(v; i, s)$ represents a general function form and is not limited to any particular distribution form (e.g. a Weibull distribution, as described in Appendix~\ref{app:parameterizations}). In practice, this means the user is free to choose any distribution model of their liking (with parameters possibly dependent on occurrence period $i$ and claim size $s$), as is the case for all other modules that follow (but with a slight modification on the dependence structure).}

The \rev{default} version of $F_{V|i,s}(v;i,s)$ is set out in Appendix~\ref{app:parameterizations}.

\subsubsection{Claim settlement date} \label{sec:module4}
Settlement delay $w_{ir}$ is the realization of a survival random variable $W_{ir}$ with survival function $\overline{F}_{W|i,s}(w;i,s)$, specified on input as a function of $w$, and possibly dependent on occurrence period $i$ and claim size $s$, and with survival function $\overline{F}_{W|i,s} = 1 - F_{W|i,s}$ for df $F_{W|i,s}$. This is sampled according to $W_{ir} = F_{W|i,s}^{-1} \left( Z_{ir}^{(W)} \right)$, where $Z_{ir}^{(W)} \sim \mathrm{Uniform}(0,1]$.

The \rev{default} version of $F_{W|i,s}(w;i,s)$ is set out in Appendix~\ref{app:parameterizations}.

\subsubsection{Number of partial payments} \label{sec:module5}
The number of partial payments $m_{ir}$ is the realization of a random variable $M_{ir}$ with df $F_{M|s} (m;s)$, specified on input as a function of $m$, and possibly dependent on claim size $s > 0$. This is sampled according to $M_{ir} = F_{M|s}^{-1} \left( Z_{ir}^{(M)} \right)$, where $Z_{ir}^{(M)} \sim \mathrm{Uniform}(0,1]$.

The \rev{default} version of $F_{M|s}(m;s)$ is set out in Appendix~\ref{app:parameterizations}.

\subsubsection{Sizes of partial payments without allowance for inflation} \label{sec:module6}
Consider claims from an Auto Liability line of business, for example. Such a claim will usually consist of:
\begin{enumerate}[label=(\arabic*)]
    \item (possibly) some small payments such as police reports, medical consultations and reports; \label{claim_type1}
    \item Some more substantial payments such as hospitalization, specialist medical procedures, equipment (e.g. prosthetics); \label{claim_type2}
    \item A final settlement with the claimant; \label{claim_type3}
    \item A smaller final payment, usually covering legal costs. \label{claim_type4}
\end{enumerate}

The settlement with the claimant will typically account for the bulk of the claim cost, and the final payment will usually increase (not necessarily linearly) with the size of the settlement.

Claims in a number of other lines of business exhibit a similar structure, albeit with possible differences in the types of payment made. For example, in Auto Collision Damage, \ref{claim_type1} might include tow-truck costs, and \ref{claim_type2} might include replacement hire car costs. Payments of type \ref{claim_type4} might be negative and account for third party recoveries and salvage.

Some claims may be trivial, involving only preliminary costs of types \ref{claim_type1} and \ref{claim_type2}. It is assumed \rev{by default} that these claims are characterized by $m_{ir} \leq 3$, and that the more substantial claims, with payments of types \ref{claim_type3} and \ref{claim_type4}, are characterized by $m_{ir} \geq 4$.

\textbf{Case $\mathbf{m_{ir} \geq 4}$.} In accordance with the above commentary, payments of types \ref{claim_type3} and \ref{claim_type4}, i.e. the last two payments in respect of the claim, are simulated first. Initially the sum of the two is simulated, and then this amount is apportioned between the two payments.

The sum of the two payments are sampled, as a proportion of claim size $s_{ir}$, according to $1 - \Big[ P_{ir}^{(m_{ir}-1)} + P_{ir}^{(m_{ir})} \Big] \sim F_{P|s}^{-1} \left( Z_{ir}^{(\overline{P})} \right)$, where $Z_{ir}^{(\overline{P})} \sim \mathrm{Uniform}(0,1]$. The \rev{default} version of $F_{P|s}$ is set out in Appendix~\ref{app:parameterizations}.

Let the proportion of the total represented by the settlement be denoted
\[
\overline{Q}_{ir} = \frac{P_{ir}^{(m_{ir} - 1)}}{P_{ir}^{(m_{ir} - 1)} + P_{ir}^{(m_{ir})}} ,
\]
which is sampled according to $\overline{Q}_{ir} \sim F_{\overline{Q}|s}^{-1} \left( Z_{ir}^{(\overline{Q})} 
\right)$, where $Z_{ir}^{(\overline{Q})} \sim \mathrm{Uniform}(0, 1]$. The \rev{default} version of $F_{\overline{Q}|s}$ is set out in Appendix~\ref{app:parameterizations}.

This leaves the proportion $1 - \left[ P_{ir}^{(m_{ir}-1)} + P_{ir}^{(m_{ir})} \right]$ of $s_{ir}$ to be accounted for by partial payments $s_{ir}^{(1)}, s_{ir}^{(2)}, \dots, s_{ir}^{(m_{ir} - 2)}$.

The proportions $p_{ir}^{(m)}, m = 1, \dots, m_{ir} - 2$ require normalization so that
\[
\sum_{m=1}^{m_{ir}-2} p_{ir}^{(m)} = 1 - \left( p_{ir}^{(m_{ir} - 1)} + p_{ir}^{(m_{ir})} \right) .
\]
They are first simulated in unnormalized form $\hat{p}_{ir}^{(m)}$ and these quantities then normalized. Thus, $\hat{p}_{ir}^{(m)}$ is the realization of a random variable $\hat{P}_{ir}^{(m)} \sim F_{\hat{P}|m}^{-1} \left( Z_{ir}^{\hat{P}|m} \right)$, where $Z_{ir}^{\hat{P}|m} \sim \mathrm{Uniform}(0,1]$. The \rev{default} form of $F_{\hat{P}|m}$ is set out in Appendix~\ref{app:parameterizations}.

Then normalization takes the form
\[
p_{ir}^{(m)} = \left[ 1 - \left( p_{ir}^{(m_{ir} - 1)} + p_{ir}^{(m_{ir})} \right) \right] \frac{\hat{p}_{ir}^{(m)}}{\hat{p}_{ir}^{(1)} + \dots + \hat{p}_{ir}^{(m_{ir} - 2)}} , \quad m = 1, \dots, m_{ir} - 2 .
\]

A summary of the construction of partial payments is as follows:
\begin{itemize}
    \item Simulate $p_{ir}^{(m_{ir} - 1)} + p_{ir}^{(m_{ir})}$.
    \item Simulate $\overline{q}_{ir}$.
    \item Calculate $p_{ir}^{(m_{ir} - 1)} = \overline{q}_{ir} \left( p_{ir}^{(m_{ir} - 1)} + p_{ir}^{(m_{ir})} \right)$, $p_{ir}^{(m_{ir})} = (1 - \overline{q}_{ir}) \left( p_{ir}^{(m_{ir} - 1)} + p_{ir}^{(m_{ir})} \right)$.
    \item Simulate $\hat{p}_{ir}^{(m)}$, $m = 1, \dots, m_{ir} - 2$.
    \item Normalize to obtain $p_{ir}^{(m)}$, $m = 1, \dots, m_{ir} - 2$.
    \item Calculate $s_{ir}^{(m)} = p_{ir}^{(m)} s_{ir}$, $m = 1, \dots, m_{ir}$.
\end{itemize}

\textbf{Case $\mathbf{m_{ir} = 2 \text{ or } 3}$.} The calculations proceed as above, except without settlement and associated partial payments. Thus,
\begin{itemize}
    \item for actual $m_{ir} = 2$, proceed as if $m_{ir} = 4$ with $p_{ir}^{(m_{ir} - 1)} + p_{ir}^{(m_{ir})} = 0$;
    \item for actual $m_{ir} = 3$, proceed as if $m_{ir} = 5$ with $p_{ir}^{(m_{ir} - 1)} + p_{ir}^{(m_{ir})} = 0$.
\end{itemize}

\textbf{Case $\mathbf{m_{ir} = 1}$.} For this degenerate case, $s_{ir}^{(1)} = s_{ir}$ of course.

\rev{The above sets up the simulation of partial payment sizes based on the number of partial payments, $m_{ir}$. While we believe this structure is generalizable to many lines of businesses, users of \texttt{SynthETIC} can easily amend this structure by feeding in an alternative simulation model.}

\subsubsection{Distribution of payments over time} \label{sec:module7}
As noted in Section~\ref{sec:notation}, the epoch of the $m$-th partial payment is $u_{ir} + v_{ir} + d_{ir}^{(1)} + \dots + d_{ir}^{(m)}$. Let this be denoted $\overline{t}_{ir}^{(m)}$. The payment period in which this payment falls is $\lceil u_{ir} + v_{ir} + d_{ir}^{(1)} + \dots + d_{ir}^{(m)} \rceil$.

The $d_{ir}^{(m)}$ require normalization so that $d_{ir}^{(1)} + \dots + d_{ir}^{(m_{ir})} = w_{ir}$, and so their simulation parallels that of the $p_{ir}^{(m)}$ just above. Thus, $\hat{d}_{ir}^{(m)}$ is the realization of a random variable $\hat{D}_{ir}^{(m)} \sim F_{\hat{D}|m}^{-1} \left( Z_{ir}^{\hat{D}|m} \right)$, where $Z_{ir}^{\hat{D}|m} \sim \mathrm{Uniform}(0,1]$. The \rev{default} form of $F_{\hat{D}|m}$ is set out in Appendix~\ref{app:parameterizations}.

Normalization takes the form
\[
d_{ir}^{(m)} = w_{ir} \frac{\hat{d}_{ir}^{(m)}}{\hat{d}_{ir}^{(1)} + \dots + \hat{d}_{ir}^{(m_{ir})}}, \quad m = 1, \dots, m_{ir} .
\]

\subsubsection{Claim inflation} \label{sec:module8}
The actual dollar value of a constant dollar partial payment is
\[
s_{ir}^{*(m)} = s_{ir}^{(m)}
\frac{ f \left( t_{ir}^{(m)} \right) }{ f(1) } 
\frac{ g_O(i \mid s_{ir} ) }{ g_O(1 \mid s_{ir}) } 
\frac{ g_C \left( t_{ir}^{(m)} \mid s_{ir} \right) }{ g_C(1 \mid s_{ir}) } .
\]

\subsubsection{Out-of-bounds transactions} \label{sec:modulez}
For the purpose of the present sub-section, a transaction includes occurrence, notification, settlement, or a payment.  Occasionally, simulated transactions will be out-of-bounds, i.e. take place beyond the end of development period $I$, where development periods are numbered $1,2,\ldots$.  

In these cases, the simulated epoch of occurrence of the transaction is maintained throughout the simulation of details of the claim concerned, other than in the exceptions noted below.  For example, if settlement occurs at development time $j>I$, and delays between partial payments depend on settlement delay, then the simulated value of $j$ will be used in the simulation of inter-partial delays.  
Only at the stage where transactions are assigned to development periods for the purpose of \reva{either addition of inflation or (optionally) tabulation} is the epoch of occurrence varied.  In these circumstances, the transaction is deemed to have occurred at the end of development period $I$. \reva{For tabulation, the user will also have the option to leave the out-of-bound transactions in a separate ``tail" cell, from which they can decide whether to aggregate them in the maximum development period or to expand the triangle.}

\reva{Accumulation into the maximum development period} will cause some concentration of transactions at the end of development period $I$.  Usually, the concentration will be small to the extent of virtual immateriality.  If the effect is material, this may indicate that the user’s selection of parameters determining transaction times is not well matched to the maximum number of development periods allowed, and consideration might be given to changing one or the other.

\subsection{Modularity} \label{sec:modularity}
\texttt{SynthETIC} has been structured so as to generate a portfolio of claims that loosely resemble those from an Auto Liability portfolio with which one of the authors is familiar. The latter is referred to here as the \textbf{``reference portfolio''}, and has been discussed in various earlier papers \citep{TaMcGr03, TaMG04, McG07b, McTaMi18}.

\rev{The structure of the default version set out in Section~\ref{sec:claim_process} and Appendix~\ref{app:parameterizations}} is quite general, and many users should be able to adopt it with changes \rev{other than in its parameters (the numerical coefficients in its algebraic structure). Slightly less simple, but still relatively easily effected, would be changes to the functional forms of sampling distributions in the default structure.} \rev{Indeed, though not always encouraged, users can also add or remove intra-model dependencies (described in Sections~\ref{sec:claim_process} and \ref{sec:data_features}) which form part of \texttt{SynthETIC}'s default architecture. Examples of such implementations are documented in the package vignette.} However, \rev{we also recognise that there remain} other cases in which some change of structure is required.

For this reason, \texttt{SynthETIC} has been coded in modular form.  Its eight modules are listed as \ref{sec:module1} to \ref{sec:module8}. The coding of any one is independent of all others so that the user may unplug any one and replace it with a version modified to his/her own purpose. \rev{\texttt{SynthETIC} includes functions to convert the replacement to a format that can be easily integrated with the later modules (see examples in the \texttt{SynthETIC} vignette).}

It is essential to note, however, that the module sequence \ref{sec:module1} to \ref{sec:module8} should be maintained. The reason is that this sequence reflects assumed dependencies whereby any specific quantities generated by any specific module may be dependent on those from prior modules. Examples are given in \ref{ex:data1} to \ref{ex:data7}.

\subsection{Data features} \label{sec:data_features}
The reference portfolio is complex, containing a number of features that can cause a certain degree of awkwardness in modelling. These are as follows.
\begin{enumerate}[label={[\thesubsection.\arabic*]}]
    \item Distribution of settlement delay depends on claim size. \label{ex:data1}
    \item Distribution of settlement delay also, conditional on claim size, varies from one occurrence period to another. \label{ex:data2}
    \item Claim sizes are subject to payment period superimposed inflation. \label{ex:data3}
    \item Rates of payment period superimposed inflation vary from one payment period to another. \label{ex:data4}
    \item Rates of payment period superimposed inflation also vary with claim size. \label{ex:data5}
    \item A legislative change occurred that affected the claim experience of subsequent occurrence periods (occurrence period superimposed inflation). \label{ex:data6}
    \item The legislative change affected claims differentially according to pre-change claim size. \label{ex:data7}
\end{enumerate}
All of these features have been incorporated in \texttt{SynthETIC}.

\section{Relation to prior literature} \label{sec:prior_lit}

\subsection{Claim simulation literature} \label{sec:claimsim_lit}
\rev{The literature contains a few predecessor simulators, namely: \citet*{CAS07}; \citet*[ASTIN Working Party on Individual Claim Development with Machine Learning]{HaGaJa17}; and \citet*{GaWu18}. These are discussed in the following sub-sections, where their major differences with \texttt{SynthETIC} are identified. A summary of those is provided in Table \ref{T_compLit}. }

\rev{\texttt{SynthETIC} is the only simulator, to our knowledge, that includes all the desirable features outlined in Section \ref{sec:desirable features}. Furthermore, it best offers a continuous spectrum of complexity, which is crucial to assess the strengths and weaknesses of a new model; see also Section \ref{sec:intro} for further discussion of this point.}

\begin{table}[H]
    \scriptsize
    \begin{tabular}{p{2cm}||p{3.6cm}|p{2.8cm}p{2.3cm}p{3.2cm}}
         & \texttt{SynthETIC} & CAS model & ASTIN WP & Gabrielli-W\"{u}thrich \\
         & (this paper) & (see Section \ref{sec:cas_model}) & (see Section \ref{astin_model}) & (see Section \ref{sec:GaWu_model}) \\ \hline 
        Availability & Open source: GitHub and CRAN & Open source & Not available & Available online \\ \hline
        Inspiration & Auto Liability portfolio, but easily adjustable to most other LoBs (via parametric, rather than structural, changes) & Auto experience in US & Not disclosed & Closely calibrated to four unspecified LoBs (suspected Liability) of an unspecified real insurance portfolio \\\hline
        Time definition & Fully flexible in terms of time length and granularity & Discrete dates & Years & Years (12) \\\hline
        Claim size (severity) distributions & Fully flexible, any distribution or GLM. Fully compatible with \texttt{actuar}.
Covariates are allowed in very flexible way

See Section \ref{sec:module2}
 & Select from prescribed set, options to modify to allow both zero claims and deductibles & Lognormal with prescribed location and dispersion parameters & Lognormal with location as a linear function of output of the prior layer of neurons.
Covariates (albeit restricted to a small prescribed set) are allowed (but their effects on the simulated data are also prescribed) and nonlinear trends can be incorporated (albeit implicit)
 \\\hline
        Notification and settlement event
         & Notification and settlement delay distributions can be specified by user, but package includes default distributions.

See Sections \ref{sec:module3}-\ref{sec:module4}
 & Automatic fit from prescribed set of distributions, or empirical distribution & Not dealt with & Notification simulated from a prescribed neural network, settlement delays sampled from a given categorical distribution
 \\\hline
 Partial payments & Explicit random number (see Section \ref{sec:module5}), with sizes also simulated according to a realistic pattern (see Section \ref{sec:module6}) & No provision & Indirectly & Deterministic payment pattern \\\hline
        Claim development & Any tail length is possible (including very long)
        
See Section \ref{sec:module7}
 & Calculated using user-specified year-to-year development factors (randomized through the standard deviation of the factors), but only the ultimate losses are returned & Shape $F_P(j)$ drawn from a lognormal with prescribed location and dispersion parameters & Relatively short-tailed \\\hline
        Claim inflation & All rates can be set arbitrarily:
        
-	Base inflation via quarterly rates

-	Superimposed inflation via both payment periods and occurrence periods

See Section \ref{sec:module8}
 & Via ``severity index'' & No explicit allowance & Not explicitly, but implicitly through trend \\\hline
        Possible model internal dependencies & Explicit modelling 
In the default version these include:

- distribution of settlement delay depends on claim size

- notification delay may depend on occurrence period and claim size

- settlement delay may depend on occurrence period and claim size

- number of partial payments may depend on claim size

- size of partial payments depend on the sizes of those preceding

Users can add or remove intra-model dependencies.
 & Via correlation only, including:
 
-	Claim size and settlement delay

-	Notification delay and settlement delay
 & Frank copula between the paid and outstanding development patterns & -	Notification delay impacts number of years with payment and claim size.
 
-	Partially across development years only, as payment pattern is deterministic
 \\\hline
        Extras & -  Modular coding for enhanced flexibility
        
- Allows for structural changes such as legislative changes
 & - Claim re-openings
 
- Generates distributional assumptions by fitting a prescribed set of distributions on a claim file provided by the user (following a specific format)
 &  & -	Claim recoveries
 
-	Claim re-openings
 
    \end{tabular}
    \caption{\rev{Comparison of existing loss reserving simulators}}
    \label{T_compLit}
\end{table}

\subsubsection{CAS Loss Simulation Model Working Party} \label{sec:cas_model}
The simulator incorporates controls on the following features:
\begin{itemize}
    \item Notification delay;
    \item Settlement delay;
    \item Claim size.
\end{itemize}
Each of these three may be specified by selecting from a prescribed set of standard distributions. There are options to modify the claim size distribution to allow for either or both of zero claims and the effect of deductibles.

The simulator is open source. It is calibrated against the Auto experience of a number of US states, although the user may over-ride this calibration.

The program includes dependencies between:
\begin{itemize}
    \item claim size and settlement delay; and
    \item notification delay and settlement delay.
\end{itemize}
In both cases, dependency is expressed in terms of correlation, which differs from the present simulator. Claim inflation is included by means of a user-defined ``severity index''.

No provision for partial payments is included.

\subsubsection{ASTIN Working Party on Individual Claim Development with Machine Learning} \label{astin_model}
This simulator generated samples of both paid and incurred amounts. The latter is not of concern for present purposes. Although the model of claim payments is expressed in terms of continuous time, it is implemented in discrete time (years).

Let $j$ denote continuous development time, with origin at the commencement of the occurrence period. Cumulative claim payments to development time $j$, denoted $P(j)$, are defined as 
$P(j) = U F_P(j)$, where $U$ is the ultimate claim size and $F_P(j)$ is the proportion of it paid by development time $j$.

Claim size $U$ is sampled from a log normal distribution with prescribed location and dispersion parameters. For each $j = 1,2, \dots$, $F_P(j)$ is drawn from a log normal distribution also with prescribed location and dispersion parameters. In particular, the location parameter at development time $j$ takes the form $\alpha \exp(-((j - \tau) / \lambda)^2)$ with the quantities $\alpha, \tau ,\lambda$ used to calibrate the payment pattern to short and long tail forms.

This simulator does not deal with notification or settlement events. It does not include explicit allowance for inflation.

Since $F_P(j)$ is a continuous function of $j$, the simulator envisages a claim as paid continuously by infinitesimal amounts. As pointed out above, however, it is implemented in discrete time, with payments occurring in each development year.

In this sense, allowance is made for partial payments. On the other hand, each claim is just a miniature version of aggregate payment experience, subject to random perturbation. This would not be a realistic representation of partial payments for most lines of business.

\subsubsection{Gabrielli-W\"{u}thrich} \label{sec:GaWu_model}
The simulator of \citet{GaWu18} generates claim payments in discrete time (years) over 12 development years. It is calibrated against a specific set of real data from four lines of business. The lines of business are not stated, though the fact that the data include ``labor sector of the injured'' and ``part of the body injured'' suggest Casualty lines, possibly Employers Liability. The code is available on-line.

The simulator \rev{generates} the following features:
\begin{itemize}
    \item Notification delay (in development years);
    \item Settlement delay (in development years);
    \item Claim size, including the proportion of zero-cost claims;
    \item Number of development years with payment;
    \item Claim payments by development year.
\end{itemize}

Less usual features included in the simulator are:
\begin{itemize}
    \item Covariates (accident year, accident quarter, line of business, labor sector of the injured, age of the injured, part of the body injured) which differentiate claims experience;
    \item Claim recoveries; and
    \item Claim re-openings.
\end{itemize}

The inclusion of accident year and quarter in covariates enables the model to accommodate (non-linear) time trends and seasonalities across accident periods in all model components. The case of claim size is of special interest in this context. This is included as a log normal variate with location parameter expressed as a linear function of the output of the prior layer of neurons.  

Hence the logarithm of claim size incorporates a non-linear trend across accident periods. The simulator does not include explicit allowance for inflation but, according to the trend just discussed, includes implicit accident year inflation. It does not include calendar period inflation.

Dependencies between some simulated quantities are included in the model. For example, the number of development years containing payment depends on the notification delay; claim size depends on notification delay and number of development years containing payment.

Dependencies between payments of different development years are included partially by virtue of the simulation of number of development years with payment. For any fixed number of these development years, a deterministic payment pattern  is assumed, whence part of the dependency between payments of different development years is excluded. The authors point out that the selection of a payment pattern could be made stochastic, but this would require careful engineering of the dependencies between development years.

The parametric structure of the simulator is fixed. The output always resembles a drawing from the data set against which the simulator is calibrated. The modeller may sometimes seek greater challenge than provided by this data set.

\texttt{SynthETIC} makes useful contributions to the already excellent simulator of \citet{GaWu18}. First, the latter appears relatively short-tailed; Table 5 of \citet{GaWu18} indicates that more than 50\% of an accident year’s cost is paid in the first development year, and roughly 90\% within the first three development years. Second, the data appear to conform reasonably well with the simple chain ladder structure.

The chain ladder will often perform well in relation to data that conform with its multiplicative cross-classified parametric structure; correspondingly, it will often perform poorly in relation to data that do not so conform.  Thus, purpose of proposed alternative models will often be to fill the gap left by the chain ladder, i.e. analysis of those data sets that are awkward for the chain ladder.  Testing of these alternative models will then require data sets that are poorly adapted to the chain ladder. \texttt{SynthETIC} was explicitly designed to facilitate such an analysis.

Similar comments can be made about the length of payment tail of a data set.  Analysis is usually most difficult in the case of long-tailed data sets, and so there is a need for the simulation of these. \texttt{SynthETIC} allows the user to easily specify arbitrarily long tails.

\subsection{Granular model literature} \label{sec:granmod_lit}
One family of models likely to thirst for testing data is that of \textbf{granular (or micro-) models}. These, by their nature, often attempt to model the minutiae of individual claims, and so are closely related to \texttt{SynthETIC}, which generates this detail.

A reasonably up-to-date summary of the literature of these models is given by \citet{DeMo19}. They are also discussed by \citet{Tay19}, where it is pointed out that the elaborateness of the granular model with its many parameters is likely to cause essential intra-model dependencies to be overlooked or replaced by bland and unrealistic assumptions. That reference gives a couple of specific examples, one of which relates to claim payments.

Section~\ref{sec:ex_intra_dep} illustrates this particular issue numerically. Section~\ref{sec:data_features} lists a number of other dependencies built into \texttt{SynthETIC}.

\section{Example implementation of \texttt{SynthETIC}} \label{sec:ex_implementation}

\subsection{Claim process components} \label{sec:ex_claims_comp}
An example simulation has been performed in accordance with the detailed specification set out in Appendix~\ref{app:parameterizations}. The generated experience covers 40 occurrence quarters, each tracked for 40 development quarters.

The principal features of the experience are similar to those of the reference portfolio, as set out in \ref{ex:data1} to \ref{ex:data7}, though slightly less extreme than that case. All of those features are present in the simulated experience. Some specific detail of relevance is the following:
\begin{enumerate}[label=\thesubsection.\arabic*.]
    \item \label{ex:imp1} Settlement delays, in addition to depending on claim size, decline gradually by 15\% over the first 20 occurrence quarters, but is stable over subsequent quarters with the exception noted in \ref{ex:imp4} 
    \item \label{ex:imp2} Base inflation occurs at 2\% per annum. 
    \item \label{ex:imp3} Payment period superimposed inflation is very high (30\% per annum) for the smallest claims, but zero for claims exceeding \$200,000 in dollar values of payment quarter 1. The rate of inflation varies linearly between claim sizes of zero and \$200,000. 
    \item \label{ex:imp4} There was a legislative change at the end of occurrence quarter 20, which, in the main, affected smaller claims in all subsequent occurrence periods. As a result, settlement delays of claims up to \$20,000 (in dollar values of payment quarter 1), already reduced by 15\% (see \ref{ex:imp1}), immediately decline by a further 20\%, but this effect is gradually eroded over the next 10 occurrence periods. At the same time, claims of up to \$50,000 reduce in size. The reduction is 40\% for the smallest claims, nil for claims of \$50,000, and the reduction varies linearly between these claim sizes.
\end{enumerate}

\begin{figure}[htbp!]
    \centering
    \includegraphics[scale=0.5]{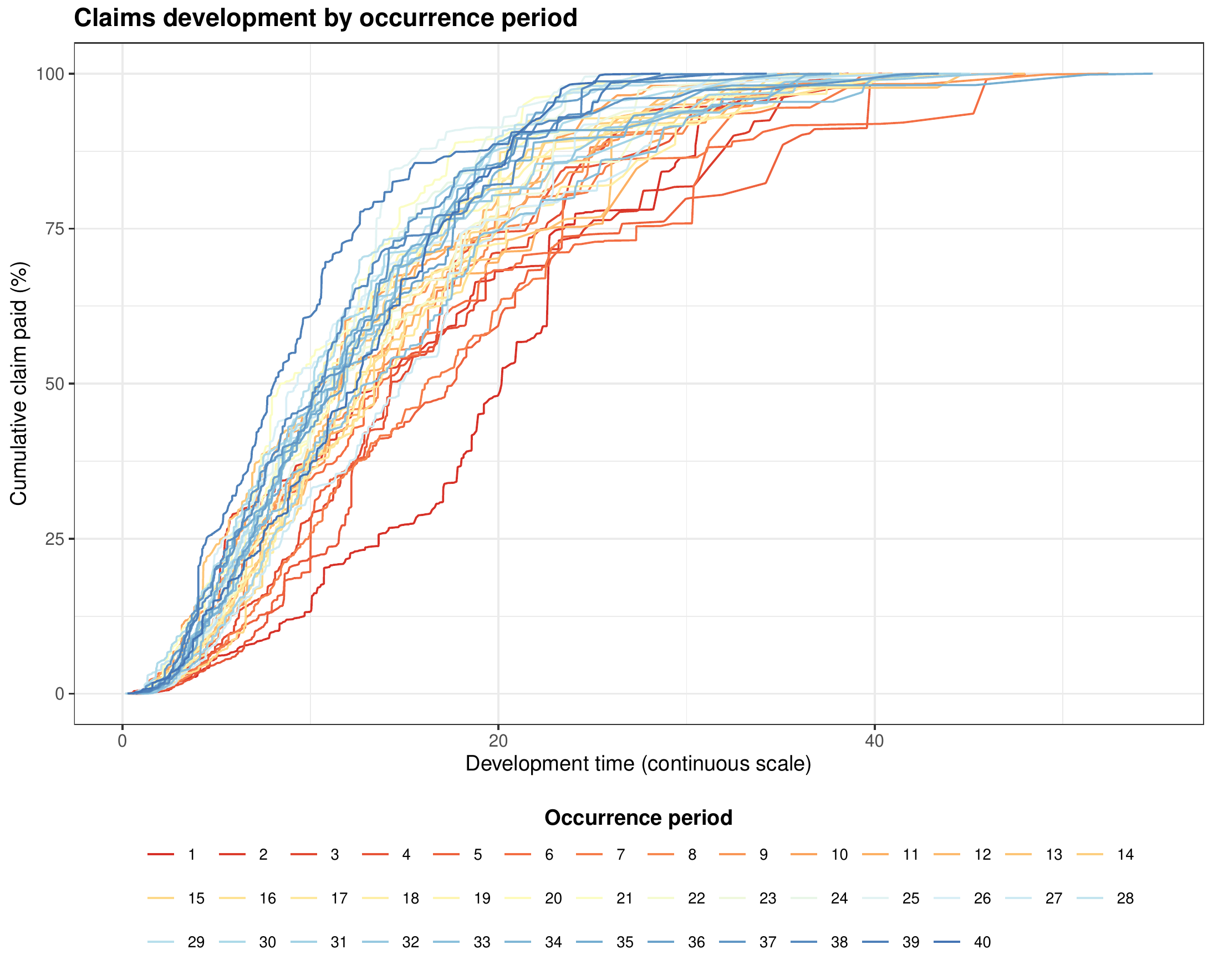}
    \caption{Development pattern of \texttt{SynthETIC}'s test implementation}
    \label{fig:claim-development}
\end{figure}

\rev{Some of these features cause a steady shortening of the claim payment pattern, which is illustrated in Figure~\ref{fig:claim-development}.} %[GT - "Figure 1" is hard coded here.  I did not know how to add it as a dynamic reference]

\subsection{Numerical results} \label{sec:ex_numerical}
\subsubsection{Comparison with chain ladder} \label{sec:ex_chainladder}
The data features set out in \ref{ex:data1} to \ref{ex:data7} and \ref{ex:imp1} to \ref{ex:imp4} create dramatic breaches of the chain ladder assumptions under which the expectation of any occurrence/development period cell is simply the product of an occurrence period parameter and a development period parameter.

This opens the chain ladder to forecast error.  This has been investigated as follows:
\begin{enumerate}[label=(\alph*)]
    \item The part of the simulated individual claim data relating to payment quarters 1 to 40 has been aggregated into a $40 \times 40$ payment triangle.
    \item The chain ladder has been applied to this triangle, and a forecast of outstanding claims up to development quarter 40 obtained (in fact, there is little claim activity beyond development quarter 40).
    \item This forecast has been compared with the ``actual'' amount of outstanding claims, simulated for payment quarters 41 to 79.
\end{enumerate}
The results are set out in Table~\ref{tab:chainladder_forecasts} where the chain ladder is seen to perform very poorly. This is perhaps unsurprising in view of the data features and the extent to which they breach chain ladder assumptions. Data sets such as this are useful for testing models that endeavour to represent data outside the scope of the chain ladder. See Section~\ref{sec:application} for further comment.

\begin{table}[htb]
    \centering
    \caption{Forecast claims costs based on synthetic data set}
    \begin{tabular}{>{\centering\arraybackslash}m{4cm}>{\centering\arraybackslash}m{3cm}>{\centering\arraybackslash}m{3cm}>{\centering\arraybackslash}m{3cm}}
        \toprule
        Occurrence quarters & \multicolumn{3}{c}{Estimated loss reserve} \\
        \cmidrule{2-4}
            & Target (simulator) & Chain ladder estimate & Ratio: chain ladder to target \\
        \midrule
            & \multicolumn{1}{c}{\$M} & \multicolumn{1}{c}{\$M} & \multicolumn{1}{c}{\%} \\
        1 to 10  & 9.2  & 9.5   & 3   \\
        11 to 20 & 26.2 & 54.6  & 109 \\
        21 to 25 & 26.1 & 68.1  & 161 \\
        26 to 30 & 61.8 & 103.4 & 67  \\
        31       & 20.5 & 22.4  & 9   \\
        32       & 24.4 & 34.8  & 43  \\
        33       & 27.8 & 35.5  & 27  \\
        34       & 30.1 & 44.5  & 48  \\
        35       & 26.4 & 53.2  & 101 \\
        36       & 36.2 & 47.8  & 32  \\
        37       & 33.0 & 66.4  & 101 \\
        38       & 43.1 & 44.9  & 4   \\
        39       & 39.0 & 39.5  & 1   \\
        40       & 46.2 & 32.4  & -30 \\
        \midrule
        \textbf{Total} & \textbf{450.0} & \textbf{657.1} & \textbf{46} \\
        \bottomrule
    \end{tabular}
    \label{tab:chainladder_forecasts}
\end{table}

\subsubsection{Intra-model dependencies} \label{sec:ex_intra_dep}
The general structure of payments in respect of a specific claim is described in Section~\ref{sec:claim_process}. Here, larger claims are envisaged as involving a settlement payment, usually shortly before closure, and with this payment ending to dominate other payments in respect of this claim. Smaller claims do not exhibit this feature.

One effect of this is that, if a large $m$-th partial payment is experienced, then the likelihood that the next payment in respect of the same claim will also be large is very much reduced at larger values of $m$. Likewise, if only small or medium payments have been observed, then the likelihood of a larger payment in the future is increased.

This stands in contrast with some granular models, which assume independence between the sizes of partial payments. This creates difficulties for the granular model, since the incorporation of the required dependencies may be awkward without the assumption of a claim payment structure such as that used in the simulator under discussion here.

The type of dependency described above is illustrated in Table~\ref{tab:dependencies}, which is an abridged summary of the simulated data set. It displays, for selected values of $m$ and various size ranges of $m$-th partial payments, the average size of the $(m+1)$-th partial payment, where inflation has been excluded throughout. The table confirms the expected dependencies between the sizes of successive partial payments.

\begin{table}[htb]
    \centering
    \caption{Simulated sizes of successive partial payments (excluding inflation)}
    \begin{tabularx}{.9\textwidth}{P{2.5cm} RRRRRR}
        \toprule
        \multirow{2}{3.5cm}{Size of the $m$-th partial payment} 
            & \multicolumn{6}{c}{Size of $(m+1)$-th partial payment for $m=$} \\
            & 1 & 2 & 3 & 5 & 7 & 9 \\
        \midrule
        \multicolumn{1}{c}{\$K} & \multicolumn{1}{c}{\$K} & \multicolumn{1}{c}{\$K} & \multicolumn{1}{c}{\$K} & \multicolumn{1}{c}{\$K} & \multicolumn{1}{c}{\$K} & \multicolumn{1}{c}{\$K} \\
        0 to 1    & 0.5       & 0.9       & 1.9  & 6.6  & 0.9   & 3.9 \\
        1 to 2    & 1.6       & 3.8       & 6.0  & 8.0  & 12.7  & 20.8 \\
        4 to 5    & 4.5       & 12.0      & 21.8 & 31.4 & 36.4  & 53.6 \\
        8 to 10   & 8.8       & 31.5      & 48.9 & 94.3 & 133.0 & 124.8 \\
        20 to 50  & 27.3      & 253.9     & 55.6 & 25.5 & 4.6   & 4.4 \\
        50 to 100 & 58.7      & 1099.0    & 8.1  & 8.0  & 8.9   & 10.1 \\
        over 100  & no claims & no claims & 32.2 & 31.8 & 33.3  & 37.8 \\
        \bottomrule
    \end{tabularx}
    \label{tab:dependencies}
\end{table}

\vspace{-4mm}

\newpage

\section{Application of \texttt{SynthETIC}} \label{sec:application}

\subsection{Data complexity} \label{sec:data_complexity}
The synthetic data set with the features described in Section~\ref{sec:ex_claims_comp} contains substantial breaches of the chain ladder assumptions, as noted in Section \ref{sec:ex_chainladder}, and, in consequence, requires complex modelling. It is, however, possible to generate simple data sets using \texttt{SynthETIC}. Indeed, such data sets may be easily rendered compatible with the chain ladder.

To do so, one simply notes that compatibility is achieved if the expected distribution of claim payments across development periods is the same for all occurrence periods. Hence, \texttt{SynthETIC} will generate chain-ladder-compatible data if:
\begin{enumerate}[label=(\alph*)]
    \item \label{cond:complexity1} all of its components \ref{sec:module1} to \ref{sec:module7} are defined to be independent of occurrence period;
    \item \label{cond:complexity2} base inflation and calendar period superimposed inflation in \ref{sec:module8} occur at a constant rate per period; and 
    \item \label{cond:complexity3} occurrence period superimposed inflation in \ref{sec:module8} must be independent of all other components, but otherwise can be arbitrary.
\end{enumerate}
Condition~\ref{cond:complexity2} is the case because it is known that, insertion of a constant rate of calendar period superimposed inflation in a claim triangle will change the estimated distribution of claim payments over development periods, but will also introduce a compensating change in accident period parameters so as to leave the estimated loss reserve invariant \citep{Tay00}.

Occurrence period superimposed inflation is directly reflected in the chain ladder row parameters, which can be arbitrary. Hence condition~\ref{cond:complexity3}.

The above demonstrates that \texttt{SynthETIC} may be used to generate very simple or very complex data sets. There is obviously a vast range of intermediate cases, and so it follows that \texttt{SynthETIC} may be used to generate a collection of data sets that provide a \textbf{spectrum of complexity}. Such a collection may be used to present a model under test with a steadily increasing challenge. This idea is used in \citet{McTaMi18}.

Different collections may be constructed with some components held constant and others subject to complex variation. In this way, one may explore the effectiveness of a contending model in the presence of different sources of complexity.

\subsection{Claim settlement dates} \label{sec:settlement_dates}
The architecture of \texttt{SynthETIC} depends heavily on claim closures. Payments in respect of claims other than small tend to be concentrated toward the settlement date in the example reported in Section~\ref{sec:ex_implementation}, because the settlement payment occurs on average three months ahead of closure (see Appendix~\ref{app:parameterizations}).

This means that claim closure date is a strong marker for the occurrence of a large payment. This tends to be realistic for Casualty lines that do not pay claims as annuities (such as Workers Compensation), and even for some major Property lines (such as Auto and Home).

Reserving models that give prominence to closure count data are discussed in \citet{Tay86, Tay00} and also in \citet{HuQiWu16}. The sequence of papers by \citet{TaMcGr03}, \citet{TaMG04}, \citet{McG07b} and \citet{McTaMi18}, based on the same data set, also used this type of model.

These models rely, of course, on accuracy of the count data, and this is sometime lacking. However, \citet{TaXu16} applied one of these models to US data from the NAIC \citep{MeSh11}, which consisted of a variety of claim count qualities. They were able to demonstrate that the use of closure counts produced tighter forecasts than obtained without it.

Despite this, the literature has largely passed over such models. Some modelling is based on both claim amounts and claim notification counts, e.g. the \textbf{double chain ladder} \citep{MMNiVe13}. Notification counts are of obvious value in enabling estimation of ultimate numbers of incurred claims, which can serve as measures of exposures over accident years.

These exposures would usually be multiplicative in the modelling of claim amounts, with expected claim payments in a triangle cell assumed proportional to number of claims incurred in the relevant accident period. This is similar to the Payments per Claim Incurred model discussed in \citet{Tay86, Tay00}.

However, notification counts are less adapted to modelling of any departure from this proportionality. In longer tailed lines, notification and settlement may be separated by substantial numbers of years, so the contribution of notification counts is likely to be oblique at best, and usually inferior to that of closure counts (assuming the latter to be reliable). \texttt{SynthETIC}, by its construction, linking claim size and partial payments to settlement delay, provides a suitable test environment for claim models based on claim settlement counts.

The ultimate form of claim data for ML models is transactional, which requires the modelling of partial payments. Relatively few contributions to the literature confront this issue, and those that do are sometimes bedevilled by unrealistic assumptions of independence of the sort discussed in Section~\ref{sec:ex_intra_dep}. \texttt{SynthETIC} includes partial payments in a realistic manner, recognizing the dependencies between them, and so provides a suitable test environment for claim models at a transactional level.

% Edited - GT
\subsection{Real data} \label{sec:real_data}
\rev{It is mentioned in Section~\ref{sec:modularity} that the default version of \texttt{SynthETIC} generates data generally compatible with a real Auto Bodily Injury portfolio. The actual data from that portfolio has been successfully modelled many times, including in the literature (see, for example, the citations in Section~\ref{sec:modularity}).}

\rev{It is to be noted, in this connection, that those analyses have usually involved operational time in 2 per cent intervals as a covariate. Although this is not the same as modelling individual claim data, it approaches that situation.}

\rev{In this sense, \texttt{SynthETIC} is seen to generate data that is essentially real, and capable of being modelled with a good deal of complexity.}

\section{\texttt{SynthETIC} repository} \label{sec:sim_repository}
\texttt{SynthETIC} is published as an open-source R package on the Comprehensive R Archive Network (CRAN) at \url{https://CRAN.R-project.org/package=SynthETIC} \citep{R-SynthETIC}. The package is licensed under GPL-3. The source code is fully available and users are free to copy, modify, or redistribute the program or any of its derivative versions. The re-distribution must not impose any further restrictions on the rights granted by the GPL.

\texttt{SynthETIC} has functions to sequentially simulate each of the eight modules as outlined in Section~\ref{sec:architecture}. The default setting assumes probability distributions of the quantities as detailed in the Appendix~\ref{app:parameterizations} (except in the case of base inflation), but the distributional assumptions can be easily modified by users to match their specific claims experiences. Users can choose to output their simulated claims in the form of a chain-ladder square by occurrence and development periods, or alternatively a structured data set at either a claim or a payment level. The data set at the payment level can then be used to visualize the claims development over time \rev{(see e.g. Figure~\ref{fig:claim-development})}. A test data set generated under the \rev{default} specification is also available as part of the package. A full demonstration of \texttt{SynthETIC}'s functionalities, \rev{including examples on alternative specifications and replacement modules,} can be accessed by running \texttt{RShowDoc("SynthETIC-demo", package = "SynthETIC")} in the R console after the installation of the package.

Users can install \texttt{SynthETIC} from the CRAN repository by running in R \texttt{install.packages("SynthETIC")}. A development version of the program is also available on \url{https://github.com/agi-lab/SynthETIC}. The GitHub repository contains, in addition to the package code, \begin{itemize}
    \item A PDF version of the package reference manual (also available on \url{https://CRAN.R-project.org/package=SynthETIC});
    \item A chain-ladder analysis of the test data set (discussed in Section~\ref{sec:ex_implementation}), in an Excel spreadsheet.
\end{itemize}

\section{Conclusions} \label{sec:conclusion}
The foregoing sections describe an individual claim simulator, newly introduced to the literature. Its code is open source, and it is modular with unpluggable and re-pluggable modules for the convenience of the user. Default modules are provided, based broadly on real data, and incorporating the data features highlighted in \ref{ex:data1} to \ref{ex:data7}.

Three existing simulators are discussed in Section~\ref{sec:claimsim_lit}. These are all useful within specific contexts, but none contains all features \ref{ex:data1} to \ref{ex:data7}. The Gabrielli-W\"{u}thrich simulator (Section~\ref{sec:GaWu_model}) is perhaps the most extensive in its allowance for dependencies between the various observations on a single claim, but it is calibrated to a single data set, and so its simulations always mirror that data set.

The simulator proposed here reflects a number of desirable (but complicated) data features, and is flexible in the form of model used to generate claim experience. Data features within its modules are easily adjusted, or the modules completely replaced. This enables the generation of a collection of data sets providing a spectrum of complexity with which to challenge a proposed model (Section~\ref{sec:data_complexity}).

\texttt{SynthETIC} may be of especial value in testing granular models. These sometimes include unrealistic assumptions of independence between different variates within the model. \texttt{SynthETIC}, on the other hand, contains built-in dependencies, e.g. between different partial payments.

\section*{Acknowledgements}
%This paper was presented at the 23rd International Congress on Insurance: Mathematics and Economics (IME) in July 2019 (Munich, Germany) and at the 54th Actuarial Research Conference (ARC) in August 2019 (Purdue University, USA). The authors are grateful for constructive comments received from colleagues who attended those events.

\rev{The authors are very grateful for comments from two referees, which led to significant improvements of the paper and \texttt{SynthETIC}, the \texttt{R} package that we developed for this work.}

This research was supported under Australian Research Council's Linkage (LP130100723) and Discovery (DP200101859) Projects funding schemes.  Melantha Wang acknowledges financial support from UNSW Australia Business School. The views expressed herein are those of the authors and are not necessarily those of the supporting organisations.

%\newpage

\bibliographystyle{elsarticle-harv}
\bibliography{libraries}

\newpage

\appendix
%\newpage
%\begin{appendices}
\section{Parametrizations} \label{app:parameterizations}
The following table displays the formal parameterization of modules \ref{sec:module1} to \ref{sec:module8} for the example of Section~\ref{sec:ex_implementation}.

\begin{center}
 \footnotesize % long table (longtablex)
\begin{tabular}{p{3cm}p{2.5cm}p{9cm}}
\toprule
{\textbf{Parameter type}} & {\textbf{Functional form}} & {\textbf{Numerical parameters}} \\
\midrule
\textbf{Global}        & & 
Time unit $= 1/4$ (calendar quarter) \newline
Reference claim size $= 200,000$$^1$   %\footnote{This component is defined in terms of claim size.  The definition here displays claim size in raw (uninflated) units.  The inputs to the example application of \texttt{SynthETIC}, on the other hand, express claim sizes as multiples of a reference claim size equal to 200,000.  For example, the amount of 100,000 that appears in the definition of claim notification delay is expressed in \texttt{SynthETIC} as $0.5\times 200,000$.}
\\
\midrule
\textbf{Claim occurrence}$^2$   %\footnote{Exposure and claim frequency are annual effective parameters.} 
& & { $ \begin{aligned}
  I &= 40 \\
  E_i &= 12000 \\
  \lambda_i &= 0.03
  \end{aligned} $ } \\
\midrule
\textbf{Claim size}$^1$       & Power-normal & $S_{ir}^{0.2} \sim \mathrm{N}(9.5, 3)$ \\
\midrule
\textbf{Claim notification}$^1$ & Weibull & Mean = $\min(3, \max(1, 2 -                                                         \frac{1}{3}\ln(s_{ir}/100000)))$ \newline
    Coefficient of variation = 70\% \\
\midrule
\textbf{Claim closure}$^1$ & Weibull & Mean = $a(i) \min(25, \max(1, 6 + 4\ln(s_{ir}/20000)))$ \newline
Coefficient of variation = 60\% \newline
where $a(i) = \max(0.85, 1-0.0075i)$, subject to the over-riding condition that, for $s_{ir} < 20000$ and $i \geq 21$, $a(i) = \min(0.85, 0.65 + 0.02(i-21))$ \\
\midrule
{\textbf{Partial payments: \newline Number}$^1$} & & For $s_{ir} \leq 7500$, $\mathrm{Prob}(M_{ir} = 1) = \mathrm{Prob}(M_{ir} = 2) = 1/2$ \newline
For $7500 \leq s_{ir} \leq 15000$, $\mathrm{Prob}(M_{ir} = 2) = 1/3, \mathrm{Prob}(M_{ir} = 3) = 2/3$ \newline
For $s_{ir} > 15000$, distribution of $M_{ir}$ is geometric, with minimum $4$ and mean = $\min(8, 4 + \ln(s_{ir} / 15000))$ \\
\midrule
{\textbf{Partial payments: \newline Amounts}$^1$} & $F_{P|s} = \mathrm{Beta}$ &
Mean = $1 - \min(0.95, 0.75 + 0.04\ln(s_{ir} / 20000))$ \newline
Coefficient of variation = 20\% \\
& $F_{\overline{Q}|s} = \mathrm{Beta}$ & Mean = 0.9 \newline
Coefficient of variation = 3\% \\
& $F_{\hat{P}|m} = \mathrm{Beta}$ & Mean = $\left( 1 - \left( p_{ir}^{(m_{ir} - 1)} + p_{ir}^{(m_{ir})} \right) \right) / \left( m_{ir} - 2 \right)$ \newline
Coefficient of variation = 10\% \\
\midrule
{\textbf{Distribution of payments over time}} & $F_{\hat{D}|m} = \mathrm{Weibull}$ &
For $m_{ir} \geq 4$ and $m = m_{ir}$, \newline
Mean = 3 months (converted to the relevant time units) \newline
Coefficient of variation = 20\% \newline
For $m_{ir} \geq 4$ and $m < m_{ir}$, or $m_{ir} < 4$, \newline
Mean = mean claim closure delay (from notification) / $m_{ir}$ \newline
Coefficient of variation = 35\% \\
\midrule
\textbf{Base inflation} & $f(\overline t)$ & $=(1+\alpha)^{\overline{t}}$, where $\alpha$ is equivalent to an increase of 2\% p.a., allowing for the relevant time units\\
\midrule
\textbf{Superimposed inflation}$^1$ & $g_O(u|s)$ & $= 1$ for $u \leq 20= 1 - 0.4\max(0, 1 - s/50000)$ for $u > 20$ \\
& $g_C(\overline{t}|s)$ & $= (1 + \beta(s))^{\overline{t}}$ with $\beta = \gamma \max(0, 1 - s/200000)$ and $\gamma$ is equivalent to a 30\% p.a. inflation rate, allowing for the relevant time units \\ 
\bottomrule
\end{tabular}
\end{center}

\noindent Notes:
\begin{enumerate}
\item This component is defined in terms of claim size.  The definition here displays claim size in raw (uninflated) units.  The inputs to the example application of \texttt{SynthETIC}, on the other hand, express claim sizes as multiples of a reference claim size equal to 200,000.  For example, the amount of 100,000 that appears in the definition of claim notification delay is expressed in \texttt{SynthETIC} as $0.5\times 200,000$.
\item Exposure and claim frequency are annual effective parameters.
\end{enumerate}

\newpage
\section{Claim development triangles of the example implementation}
\rev{For space considerations, below we show only the claim development triangles on a \textbf{yearly} basis; however, the underlying data is calculated based on quarterly development pattern and is available on the \texttt{SynthETIC} repository (see Section~\ref{sec:sim_repository}).}

\begin{table}[htbp!]
    \centering
    \footnotesize
    \begin{tabular}{crrrrrrrrrr}
    \toprule
    & \multicolumn{10}{c}{Development Year} \\ \midrule
    Accident Year & 1      & 2      & 3      & 4      & 5      & 6      & 7     & 8     & 9     & 10    \\
    \midrule
    1 & 1,066 & 9,452 & 18,927 & 30,287 & 42,525 & 56,468 & 62,876 & 69,553 & 73,852 & \textbf{76,152} \\
    2 & 1,463 & 9,836 & 25,669 & 37,002 & 46,925 & 57,862 & 64,484 & 67,626 & \textbf{74,977} & 78,175 \\
    3 & 3,021 & 15,491 & 34,694 & 49,257 & 59,293 & 66,771 & 72,653 & \textbf{78,445} & 81,514 & 82,516 \\
    4 & 2,247 & 14,847 & 33,546 & 50,969 & 60,592 & 74,500 & \textbf{79,288} & 81,996 & 83,271 & 85,974 \\
    5 & 1,892 & 20,421 & 41,594 & 62,943 & 77,215 & \textbf{87,601} & 92,755 & 96,667 & 97,922 & 101,723 \\
    6 & 4,335 & 22,748 & 47,748 & 67,447 & \textbf{84,501} & 94,869 & 100,668 & 104,826 & 106,442 & 107,307 \\
    7 & 2,015 & 21,542 & 45,726 & \textbf{67,976} & 81,616 & 91,775 & 99,333 & 104,418 & 106,847 & 106,901 \\
    8 & 4,389 & 27,606 & \textbf{55,346} & 83,075 & 99,509 & 109,076 & 115,446 & 119,931 & 122,561 & 124,959 \\
    9 & 2,955 & \textbf{32,372} & 63,773 & 92,863 & 114,928 & 131,745 & 138,791 & 145,341 & 148,514 & 152,235 \\
    10 & \textbf{4,896} & 43,386 & 78,378 & 113,860 & 133,274 & 150,229 & 162,189 & 164,782 & 165,262 & 166,216 \\ \bottomrule
    \end{tabular}
    \caption{Cumulative claim development triangle of simulated claims (\$000)}
    \label{tab:simulated triangle}
\end{table}

\begin{table}[htbp!]
    \centering
    \footnotesize
    \begin{tabular}{crrrrrrrrrr}
    \toprule
    & \multicolumn{10}{c}{Development Year} \\ \midrule
    Accident Year & 1      & 2      & 3      & 4      & 5      & 6      & 7     & 8     & 9     & 10    \\
    \midrule
    1 & 1,066 & 9,452 & 18,927 & 30,287 & 42,525 & 56,468 & 62,876 & 69,553 & 73,852 & \textbf{76,152} \\
    2 & 1,463 & 9,836 & 25,669 & 37,002 & 46,925 & 57,862 & 64,484 & 67,626 & \textbf{74,977} & 78,512 \\
    3 & 3,021 & 15,491 & 34,694 & 49,257 & 59,293 & 66,771 & 72,653 & \textbf{78,445} & 84,408 & 88,368 \\
    4 & 2,247 & 14,847 & 33,546 & 50,969 & 60,592 & 74,500 & \textbf{79,288} & 85,157 & 91,730 & 96,196 \\
    5 & 1,892 & 20,421 & 41,594 & 62,943 & 77,215 & \textbf{87,601} & 96,900 & 104,228 & 112,172 & 117,358 \\
    6 & 4,335 & 22,748 & 47,748 & 67,447 & \textbf{84,501} & 100,938 & 111,662 & 119,985 & 129,126 & 135,184 \\
    7 & 2,015 & 21,542 & 45,726 & \textbf{67,976} & 84,535 & 100,969 & 111,681 & 120,014 & 129,151 & 135,200 \\
    8 & 4,389 & 27,606 & \textbf{55,346} & 81,449 & 101,431 & 121,211 & 134,165 & 144,180 & 155,198 & 162,506 \\
    9 & 2,955 & \textbf{32,372} & 71,595 & 105,756 & 131,734 & 157,444 & 174,346 & 187,294 & 201,656 & 211,328 \\
    10 & \textbf{4,896} & 31,943 & 66,096 & 95,894 & 118,288 & 140,843 & 155,004 & 166,841 & 179,200 & 186,893 \\ \bottomrule
    \end{tabular}
    \caption{Cumulative claim development triangle predicted by the chain ladder model (\$000)}
    \label{tab:CL predictions}
\end{table}

%\end{appendices}

\end{document}